\documentstyle[prl,aps,graphicx,float]{revtex}

\begin{document}

\draft

\preprint{}

\def\be{\begin{equation}}

\def\ee{\end{equation}}

\def\bea{\begin{eqnarray}}

\def\eea{\end{eqnarray}}


\title{How backscattering off a point impurity
can enhance the current and make the conductance greater
than $e^2/h$ per channel}

\author{D.E.  Feldman$^{1,2}$ and Yuval Gefen$^3$}

\address{ $^1$Materials Science Division, Argonne National Laboratory,
9700 South Cass Avenue, Argonne, IL 60439\\
$^2$Landau Institute for Theoretical Physics, 142432
Chernogolovka, Moscow region, Russia \\ $^3$Condensed Matter
Physics Department, Weizmann Institute of Science, 76100 Rehovot,
Israel}

\maketitle

\begin{abstract}
It is well known that while forward scattering has no effect on
the conductance of one-dimensional systems, backscattering off a
static impurity suppresses the current. We study the effect of a
time-dependent point impurity on the conductance of a one-channel
quantum wire. At strong repulsive interaction (Luttinger liquid
parameter $g<1/2$), backscattering renders the linear conductance
greater than its value $e^2/h$ in the absence of the impurity. A
possible experimental realization of our model is a constricted
quantum wire or a constricted Hall bar at fractional filling
factors $\nu=1/(2n+1)$ with a time-dependent voltage at the
constriction.

\end{abstract}

\section{Introduction}

There are several motivations to study one-dimensional (1D)
quantum conductors. Quantum wires are expected to be an essential
component of future nanoelectronic devices. The analogy between 1D
electron liquid and edge states of the 2D electron gas is
conducive to the understanding of the Quantum Hall effect
\cite{QHE}. There are also many other related systems such as
vortex lines in type-II superconductors \cite{vortex}. From the
theoretical point of view 1D conductors are the simplest
non-Fermi-liquid systems. Probably, the most appealing consequence
of non-Fermi-liquid behavior is the existence of fractionally
charged quasiparticles, recently observed in experiments on
Quantum Hall systems \cite{fc}. Since the experimental setup was
based on a realization of 1D quantum wire with an impurity, the
latter problem has received considerable renewed attention.

The effect of an impurity on a gas of noninteracting electrons is evident.
It leads to backscattering, hence to suppression of the current.
Qualitatively the same happens in the case of a static impurity
in the presence of electron-electron
interaction too.
However, in that case the effect turns out to be counterintuitively strong. Even
an arbitrarily weak impurity renders the conductance of a long wire
equal to zero, thus effectively cutting the wire into two independent pieces
\cite{ML,KF,FO}. This is yet another manifestation of strong correlations in a
non-Fermi-liquid state.

While recent activity has focused primarily on the problem of
static impurities, it is also interesting to understand what
happens if the impurity potential depends on time. This question
touches upon the problem of pumping \cite{pumping} and the effect
of phonon pulses in 1D conductors \cite{phonons}. Recent works
\cite{fl} consider the effect of a time-dependent impurity on
Fermi-liquid states. Much less attention was devoted to the
interplay between a time-dependent impurity and non-Fermi-liquid
effects \cite{Chamon}. In this article we study the simplest
question of this type: how a weak,  point-like impurity whose
potential depends on time affects the conductance of a quantum
wire with a repulsive electron-electron interaction. In the
absence of interaction the answer would be obvious. The impurity
would decrease the current, the suppression of the current
depending on the strength of the impurity potential. Surprisingly,
for interacting systems the time-dependent backscattering impurity
{\it can enhance} the conductance. This is the main result of the
present Letter. Such enhancement takes place as the interaction
strength exceeds a threshold value. In terms of the Luttinger
liquid parameter $g$ the threshold is $g=1/2$. We would like to
emphasize that the predicted current enhancement is a linear
effect, and the linear conductance of a one-channel wire becomes
greater than the conductance quantum $e^2/h$ for strong repulsive
interactions.

The paper is organized as follows: first we discuss the origin of
the effect qualitatively. In the third section we formulate our
model and discuss the details of the set-up (e.g. the way how the
voltage is applied). In the forth section  details of calculations
are discussed. In conclusion we discuss the results and possible
experimental realizations.


\section{Qualitative discussion}

Let us first discuss the origin of the effect qualitatively.
While a purely qualitative discussion is insufficient to explain
the threshold value $g=1/2$, we show below
that our effect can be derived heuristically from a simple
analysis of the structure of the Hamiltonian. Our detailed
analysis yields quantitative results
concerning the weak impurity limit and is
based on the bosonization technique \cite{bosonization}.

Using the analogy between a quantum wire and edge states
of the 2D electron gas \cite{revWen},
we interpret
backscattering off a weak impurity as
tunneling between two chiral systems of right- and left-movers.
The tunneling density of states diverges as the energy approaches
the Fermi energy \cite{chiral}. In other words, backscattering
is enhanced in the two following cases:
1) the energy of the incident particle is close to the Fermi energy
of the electrons moving in the same direction;
2) the energy of the backscattered particle is close to the Fermi energy
of the electrons which move in the direction opposite to
that of the incident particle. The left and right Fermi energies
differ by the applied voltage $V$.
These statements about the dependence of the backscattering amplitude on the
energy  do not hold for noninteracting electrons.
In that case one should only note
that scattering
to an occupied state is impossible. The energy dependence
is more pronounced for the stronger  electron-electron interaction.

Now we are in a position
to consider a time-dependent impurity.
Let us assume for simplicity that the time-dependent potential is harmonic,
 $W(t)\sim U\cos\omega t$.
Note, however, that the current enhancement is possible for a
general periodic potential. In our qualitative discussion we
consider the case $\hbar\omega>eV;\hbar\omega\approx eV$, where
$V$ is the voltage and  $e= -|e|$ -- the electron charge. In this
case we predict particularly strong current enhancement. We assume
that the Fermi energy of the particles moving to the left is
$E_{\rm FL}=eV$ and the Fermi energy of the right-movers --
$E_{\rm FR}=0$. The scattering is inelastic. For small $U$ only
processes involving an emission (absorption) of a single quantum
are allowed, hence the energy change is $\pm\hbar\omega$. There
are four backscattering processes (Fig. 1) which we denote as
$L^{\pm}, R^{\pm}$, where the letter shows the type of particles
(left- or right-movers) and the $+/-$ sign corresponds to
increasing/decreasing the energy of the particle. The processes
$R^-$ and $L^-$ are suppressed by the Pauli principle since they
lead to the scattering into states which are below the Fermi
energies, $E_{\rm FL}$ and $E_{\rm FR}$ respectively. The
processes $L^+$ and $R^+$ drive particles into states above the
Fermi energy. However, only $R^+$ processes lead to the scattering
of the particles with initial energy around $E_{\rm FR}$ to final
states with energy close to $E_{\rm FL}$. As discussed above, for
such particles the probability of backscattering is enhanced. For
$L^+$ processes the particles with the initial energy close to
$E_{\rm FL}$ are backscattered into the states whose energies are
not close to $E_{\rm FR}$. Hence, the backscattering amplitude for
such processes is less than for $R^+$ processes.
 On the other hand, $L(R)^+$ processes are effective
for particles with energies $E$ in the interval $E_{\rm
FL(FR)}>E>E_{\rm FR(FL)}-\hbar\omega$, where $E_{\rm FL}>E_{\rm
FR}$ (cf. Fig. 1). Hence the number of left-moving particles which
potentially can be subject to strong $L^+$ backscattering
processes exceeds the number of right-movers which potentially are
subject to $R^+$ processes \cite{clear}. As the electron-electron
interaction increases, $R^+$ becomes more important. At some
threshold the latter begins to provide the main contribution to
the backscattering current, hence it determines its direction.
Since this process modifies right-movers to left-movers, such
backscattering enhances the current. Thus, paradoxically, the
scattering of particles backwards adds electrons to the forward
flow.

The above qualitative discussion was based on an effective
single-particle picture. Certainly, our systematic approach that
leads to essentially the same conclusions does not rely on such a
simplified picture.

Two important ingredients of the above arguments are inelastic
backscattering and energy dependence of the backscattering
amplitude. Both conditions can be realized in a non-interacting
system. This will not however result in current enhancement since
the third key ingredient is missing: Backscattering amplitudes in
a non-interacting system depend only on the absolute value of the
electron momentum but not on its direction in the case of
symmetric geometry considered in this paper. On the other hand, in
the absence of the symmetry in a non-interacting system, one gets
generation of the photocurrent instead of enhancement of the
injected current. The direction of the photocurrent is not related
to the sign of the voltage and it can flow at zero voltage.

\section{Model and sketch of derivation}

Below we provide a quantitative theory (valid for general
$V,\omega$).


We consider spinless electrons \cite{foot}
and concentrate first on the case of zero temperature.
Our starting point is the Tomonaga-Luttinger model \cite{KF,bosonization}
with a point impurity. The Hamiltonian reads

\bea \label{1} H=\int dx [-\hbar v_F(\psi_R^\dagger(x)
i\partial_x\psi_R(x)- \psi_L^\dagger(x) i\partial_x\psi_L(x) ) +
K(x)(\psi^\dagger_R(x)\psi_R(x)+\psi^\dagger_L(x)\psi_L(x))^2
\nonumber \\
+\delta(x)W(t)(\psi^\dagger_L(0)\psi_R(0)+\psi^\dagger_R(0)\psi_L(0))],
\eea where $\psi_R$ and $\psi_L$ are the fermionic field operators
of the right- and left-moving electrons, $v_F$ is the Fermi
velocity, $W(t)$ the potential of the impurity located at the
origin and $K(x)$ -- the interaction strength. We take $K(x)=K$ in
the region $-L/2<x<L/2$, and $K=0$ at $|x|\gg L$. Thus, we assume
that the electron-electron interaction is completely screened at
large $x$. Below we set $v_F$ and $\hbar$ equal to unity. We
assume that the right- and left-movers are injected from the leads
with chemical potentials $\mu_R=0;\mu_L=eV>0$. Note that in our
notation $\mu_L$ is located on the right (Fig. 1).

We use a set-up in which the interaction strength $K=0$ in the
right and left parts of the system. This is the model used in Ref.
\cite{no_g}. There are many approaches to model leads,  the bias
voltage, and other details of the systems. They reflect different
possible set-ups. Some details of the behavior may be sensitive to
the details of the set-up. We limit our discussion by the set-up
used in Ref. \cite{no_g}.

The choice of the Hamiltonian in the form (\ref{1}) assumes that
the interaction is short range. This means that the Coulomb
interaction between electrons is screened (by the gates)
\cite{footvolt}.

We follow the technical procedure developed in Ref. \cite{Wen} for
a static impurity. We employ the Keldysh formalism \cite{keldysh}.
At $t=-\infty$ the impurity potential is absent, $W(t)=0$, and
then it is gradually turned on. Thus, at the initial moment of
time the Hamiltonian (\ref{1}) is time-independent and commutes
with the operators of the numbers of right- and left-moving
particles $\hat N_R$ and $\hat N_L$. Hence, initially the system
can be described by the partition function with two chemical
potentials $\mu_R=0$ and $\mu_L=eV$ conjugated with the particle
numbers $N_R$ and $N_L$.

It is convenient to switch to the interaction representation
$H\rightarrow H-\mu_R N_R -\mu_L N_L$.  This transformation
introduces time dependence into the $\psi_L$ operator. Thus, we
have to substitute $\psi_L\rightarrow \psi_L \exp(-i eV t)$,
$\psi_L^\dagger\rightarrow\psi_L^\dagger\exp(ieVt)$ in every
expression where those operators enter. In particular, the
impurity contribution in Hamiltonian (\ref{1}) now reads
$\delta(x)W(t)(\psi^\dagger_L(0)\psi_R(0)\exp(i\omega_0t)+ h.c.)$,
where $\omega_0\equiv eV$.

Next, we derive an expression for the current operator. The
current includes the background contribution from the particles
injected from the leads and the backscattering contribution
associated with the impurity and proportional to the
backscattering rate. The background contribution is equal to the
current in the absence of the impurity. Since $\mu_L>0$, the
background particle current (of particles with $e<0$) flows to the
left.

The background current is simply $I_0=e^2V/h$ \cite{no_g}. The
backscattering contribution is time-dependent due to the
non-stationary nature of our problem. The dc contribution to the
total current must be independent of the coordinates due to the
charge conservation. Hence, it can be represented in two
equivalent forms:

\be \label{idc1} I_{dc}=I_L^\leftarrow-I_L^\rightarrow=
I_R^\leftarrow-I_R^\rightarrow, \ee where $I_L^\rightarrow$ and
$I_R^\leftarrow$ are the currents of right- and left-movers
injected form the left and right leads respectively,
$I_L^\leftarrow$ and $I_R^\rightarrow$ are the currents of left-
and right-movers incident to the left and right leads
respectively. The difference $I_L^\leftarrow-I_R^\leftarrow=
\overline{dN_L/dt}$, where the bar denotes the time average,
$dN_L/dt$ is the change in the number of left-movers in the wire
due to backscattering off the impurity. Hence,
$I_{dc}=I_R^\leftarrow-I_L^\rightarrow + \overline{dN_L/dt}$. The
currents of injected particles $I_R^\leftarrow$ and
$I_L^\rightarrow$ are the same in the absence and in the presence
of the impurity. Finally, we obtain for the  dc current

\be \label{idc2} I_{dc}=\frac{e^2V}{h}+e\overline\frac{dN_L}{dt}.
\ee

Thus, the backscattering particle current can be defined as $\hat
I_{\rm bs}=d\hat N_L/dt = - d\hat N_R/dt$, where $\hat N_L$ and
$\hat N_R$ are the particle number operators. A positive value of
the backscattering current corresponds to the enhancement of the
background current. In terms of the $\psi$-fields the
backscattering current operator (in the interaction
representation) is given by the equation

\be
\label{2}
\hat I_{\rm bs}=-iW(t)(\psi^\dagger_L(0)\psi_R(0)\exp(i\omega_0t)-
h.c.).
\ee

Since $W(t)\sim\cos \omega t$, one finds that there are two types
of time-dependent terms in the Hamiltonian and the current
operator: (i) terms proportional to $\exp(\pm
i(\omega+\omega_0)t)$ and (ii) terms proportional to $\exp(\pm
i(\omega-\omega_0)t)$. If only terms of the first (second) type
existed our problem would be equivalent to a static impurity in
the presence of an external voltage drop
$V_{1(2)}=(\omega_0\pm\omega)/e$ (in the appropriate interaction
representation). Hence, the backscattering current can be
represented as $I_{\rm bs}=I_1+I_2+I_{12}$, where $I_{1,2}$
denotes the backscattering current in the static problem with the
voltage $V_{1,2}$, and $I_{12}$ is the 'interference'
contribution. We will see that to the lowest order in the impurity
potential the interference current $I_{12}$ is ac, hence the
dc-current is made of $I_1$ and $I_2$ only. Hence if one is
interested in the averages over time $t>1/\omega$, the
contribution $I_{12}$ drops out. We know from Ref. \cite{KF} that
$I_{1,2}\sim |V_{1,2}|^{2g-1}$, where $g=\sqrt{\pi\hbar
v_F/(\pi\hbar v_F + 2K)}$ is the standard dimensionless parameter
of the Luttinger liquid. For $g<1/2$ the exponent $2g-1$ is
negative, hence the main contribution to $I_{\rm bs}$ is $I_2$.
Thus the direction of the backscattering current is the same as in
the static problem with the voltage $V_2$. At $\omega>\omega_0$
the sign of this voltage is opposite to the sign of the applied
voltage $V$. This shows that the backscattering current {\it
enhances} the background one.

\section{Quantitative analysis}

In order to actually calculate the currents
$I_{\rm dc}=I_1+I_2$ and $I_{\rm ac}=I_{12}$
we employ the bosonization transformation
\cite{bosonization,Wen}. This leads to the action

\be
\label{3}
L=\int dt dx \biggl[\frac{1}{8\pi}
((\partial_t \hat\Phi)^2-(\partial_x \hat\Phi)^2)
-\delta(x)W(t)(e^{i\sqrt{g}\hat\Phi(t,x=0)}e^{i\omega_0 t} + h.c.)\biggr],
\ee
where $W(t)=U\cos\omega t$ and the bosonic
 field $\hat\Phi$ is related
to the charge density as $\hat\rho=e\sqrt{g}\partial_x\hat\Phi/(2\pi)$.
The same action describes a quantum Hall bar with a
(time-dependent) constriction \cite{revWen}. In that
case $g$ is the filling factor $g=1/(2n+1)$.
The current operator reads

\be
\label{4}
\hat I_{\rm bs}=-iW(t)e^{i\sqrt{g}\hat\Phi(t,x=0)}e^{i\omega_0 t}+ h.c.
\ee

To find the backscattering current at the moment $t$ we
have to calculate the average

\be \label{5} \langle\hat I_{\rm bs}(t) \rangle= \langle
0|S(-\infty;t) \hat I_{\rm bs} (t) S(t;-\infty)|0\rangle, \ee
where $\langle 0|$ denotes the initial state, $S$ is the
scattering matrix. As we remember from the previous section, the
initial state is defined in terms of the two chemical potentials
of the right- and left-movers. To the lowest order in $W$

\be
\label{6}
S(t;-\infty)=1-i\int_{-\infty}^t dt'
[W(t')e^{i\sqrt{g}\hat\Phi(t',x=0)}e^{i\omega_0 t'}+ h.c.];
S(-\infty;t)=S^*(t;-\infty).
\ee

Further calculations follow the standard route \cite{Wen} and show
that the current $I=I_{\rm dc}+I_{\rm ac}$ includes a
dc-contribution $I_{\rm dc}$ and an ac-contribution $I_{\rm ac}$
of frequency $2\omega$. One evaluates expression (\ref{5}) using
the Green function of the Bose field:

\be\label{correl} \langle 0|\Phi(t)\Phi(0)
|0\rangle=-2\ln(\delta+it), \ee where $\delta$ is infinitesimal.
We first discuss the more interesting dc-contribution. It has
different forms at $\omega>\omega_0=eV$ and $\omega<\omega_0$ (cf.
Fig. 2) \bea \label{7} I_{\rm
dc}=\frac{U^2}{2(v_F\tau_c)^2}\Gamma(1-2g)\tau_c^{2g}\sin (2\pi g)
[(\omega-\omega_0)^{2g-1}-(\omega+\omega_0)^{2g-1}], ~~~
\omega>\omega_0;\\
\label{8}
I_{\rm dc}=-\frac{U^2}{2(v_F\tau_c)^2}\Gamma(1-2g)\tau_c^{2g}\sin (2\pi g )
[(\omega_0-\omega)^{2g-1}+(\omega+\omega_0)^{2g-1}], ~~~
\omega<\omega_0,
\eea
where $\tau_c\sim 1/E_{\rm Fermi}$
is a short-time cutoff.  While at $g>1/2$ the current
$I_{\rm dc}<0$ in both cases, at
strong interaction, $g<1/2$, the backscattering current becomes
positive as $\omega>\omega_0$. In other words,
in the latter case it flows in the direction of the background current.
In the limit $\omega_0\rightarrow 0$ one finds
a correction to the conductance (which would be equal to $G=e^2/h$
in the absence of the impurity \cite{no_g}). The correction
\be
\label{9}
\Delta G = \frac{e^2}{\hbar^3}\left(\frac{U}{v_F\tau_c}\right)^2
(1-2g)\omega^{2g-2}\Gamma(1-2g)\tau_c^{2g}\sin (2\pi g)
\ee
is positive for $g<1/2$. The generalization to the case when the time-dependent potential $W(t)$
contains several harmonics is straightforward.

The results (\ref{7},\ref{8}) are obtained at zero temperature. As
$T>|\omega-\omega_0|$ the expression for the current is modified.
The effect of finite temperature can be determined with the
Keldysh technique \cite{keldysh}. To avoid cumbersome expressions
we discuss here only the limiting cases. A full expression for the
current has the opposite sign and the same absolute value as the
sum of two expressions of the form (A5) \cite{KF} with
$a(t)=(\omega_0\pm\omega)t$ and $g\rightarrow 1/g$. At low
temperatures $\omega+\omega_0 > T > |\omega-\omega_0|$ the
backscattering current is given as the sum of two terms,
proportional to $(\omega-\omega_0)T^{2g-2}$ and to
$(\omega+\omega_0)^{2g-1}$ respectively. The current becomes
positive (enhancement of the total current) for $\omega>\omega_0$,
$T<(\omega+\omega_0)\left(\frac{\omega-\omega_0}{\omega+\omega_0}\right)^{1/(2-2g)}$.
As $T\gg\omega, V$ the dependence of the backscattering current on
$\omega$ drops out. The latter turns out to be always negative:
$I_{\rm bs}\sim - \omega_0 T^{2g-2}$.

There is also an ac contribution of frequency $2\omega$ to the current.
At $T=0$ it reads as follows

\bea
\label{7ac}
I_{\rm ac}(t)=\left(\frac{U}{v_F\tau_c}\right)^2
\Gamma(1-2g)\tau_c^{2g}\cos (\pi g + 2\omega t) \sin (\pi g)
[(\omega-\omega_0)^{2g-1}-(\omega+\omega_0)^{2g-1}], 
\omega>\omega_0;\\
\label{8ac}
I_{\rm ac}(t)=-\left(\frac{U}{v_F\tau_c}\right)^2
\Gamma(1-2g)\tau_c^{2g}\sin (\pi g )
[\cos(2\omega t - \pi g)(\omega_0-\omega)^{2g-1}+
\cos(2\omega t + \pi g)(\omega+\omega_0)^{2g-1}], 
\omega<\omega_0.
\eea

In our model we neglect forward scattering at the impurity. This
approximation is valid in the case of a weak impurity. Indeed,
both forward and backscattering terms in the Hamiltonian have
order $U$. The backscattering current includes three
contributions: one goes solely from backscattering, the second
solely form the forward scattering term, the third is the
interference contribution. The second contribution is zero:
forward scattering alone cannot modify the current. The
interference contribution is zero up to the second order in $U$.
Indeed, the backscattering current operator $d\hat N_L/dt=i[\hat
H, \hat N_L ]/\hbar$ has the same form (\ref{2}) both in the
presence and in the absence of forward scattering. Hence, the
interference contribution can be found from Eq. (\ref{5}). For
this one has to add a forward scattering term to the expansion of
the $S$-matrix up to the first power of $U$ Eq. (\ref{6}). A
simple calculation shows that the resulting interference
correction  is zero indeed up to the order $U^2$. Nonzero
corrections are possible in higher orders.

Thus, our calculations are restricted by the case of a weak
impurity. Note that the current enhancement is impossible for a
strong impurity: An infinite barrier cuts the systems into two
independent pieces and the total current is just zero. Hence,
there is a critical impurity strength at which our effect
disappears.

\section{Conclusion}

A possible experimental realization of our system is a one
dimensional wire in the presence of a time-dependent gate voltage
that allows to obtain a time-dependent constriction in one point.
External magnetic field must polarize electrons in the wire.

Another possible experimental realization of our model is a Hall
bar with a constriction \cite{webb}. The role of the
backscattering impurity is played by the constriction that gives
rise to weak tunneling of quasiparticles between the two edges.
The tunneling amplitude can then be made time-dependent by
application of a time-dependent gate voltage. If the system is
tuned such that it is close to a resonance \cite{KF} then it can
be described by the Lagrangian (\ref{3}) with $W(t)=U\cos\omega
t$. In the absence of tuning a static contribution $W_0$ should be
added to $W(t)$. Still if the inter-edge tunneling is weak and
$\omega$ is greater than the product of the voltage difference
between the edges and the quasiparticle charge, we expect an
enhancement of the Hall current in the FQHE (at filling factors
$g=1/(2n+1)$) as compared with the absence of a time-dependent
perturbation both at $W_0>W$ and $W_0<W$. On the other hand, the
time-dependent tunneling decreases the current in IQHE where the
filling factor $g=1$.

Both in IQHE and FQHE cases our system can be interpreted as a
rectifier: It transforms an ac gate voltage into a dc current.
There is also a relation between our problem and pumping
\cite{pumping}. Pumping requires two time-dependent parameters
while in our problem there is only one such parameter, the
impurity strength. However, as we have discussed, for the
calculation of the backscattering current the voltage can be
gauged out giving rise to an additional time-dependent parameter.
Thus, the backscattering current can be interpreted as a pumped
current. There is also an analogy between our problem and photon
assisted current: in both cases there is a time-dependent
parameter and the left-right symmetry is broken (in our case by
the voltage).

A more general question concerns the effect of external noise on
the quantum wire. One can consider the tunneling amplitude
$W(t)=\int d\omega [{U'}_\omega \cos\omega
t+{U''}_\omega\sin\omega t]$, where $\langle
({U'}_\omega)^2\rangle= \langle ({U''}_\omega)^2\rangle=
S(\omega)$ is the spectral function of the noise. The simplest
realization of a random $W(t)$ are thermal fluctuations of the
gate voltage in the set-up discussed in the previous paragraphs.
Another related problem is the effect of the irradiation by
phonons. If a hot spot is created in the region of the
constriction then phonons give rise to a time-dependent
backscattering. The total backscattering current can be obtained
from Eqs. (\ref{7},\ref{8}) integrating over $\omega$ and
substituting $2S(\omega)$ for $U^2$. Note that for white noise,
$S(\omega)={\rm const}$, the backscattering current calculated in
such way vanishes at $g<1/2$. Note however, that a mathematically
ideal white noise includes frequencies which are higher than the
Fermi energy. At such frequencies an approach based on the
Luttinger model cannot be used. This will result in small non-zero
corrections to the conductance.

In conclusion, we have found that backscattering off a point impurity
can increase the conductance of a quantum wire. This is a manifestation
of the strong electron-electron interaction
as the Luttinger liquid parameter $g<1/2$.

\acknowledgements

We thank B.L. Altshuler, A.M. Finkelstein, B.I. Halperin, Y.
Levinson and Y. Oreg for useful discussions. This research was
supported by the US DOE Office of Science under contract No.
W31-109-ENG-38, RFBR grant 00-02-17763,  GIF foundation, the
US-Israel Bilateral Foundation, the ISF of the Israel Academy
(Center of Excellence), and by the DIP Foundation.

\begin{figure} \label{fig1}
  \hfill
  \includegraphics[width=4.9in]{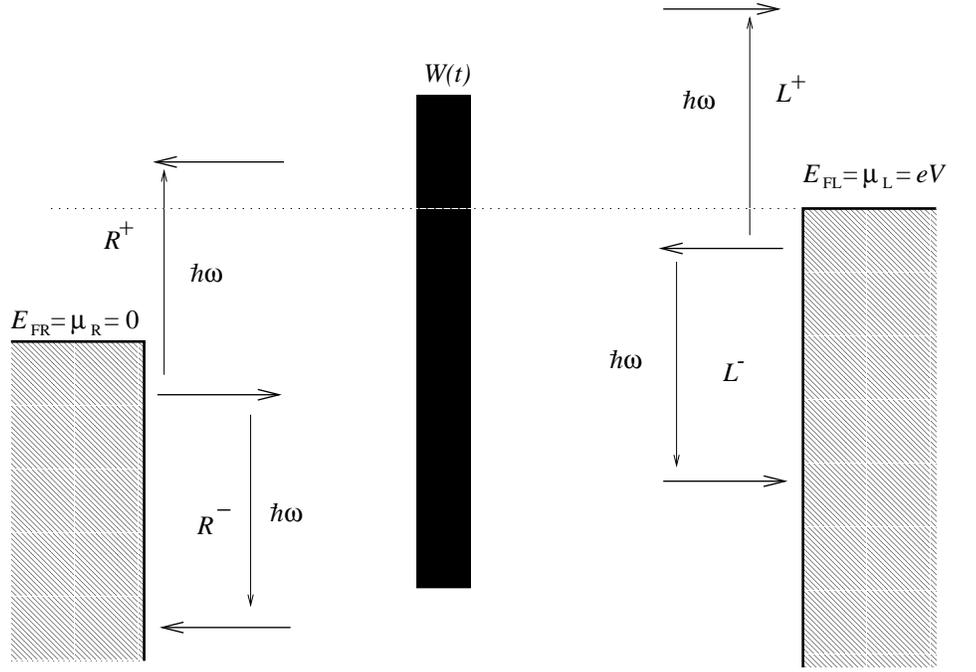}

  \hfill\hfill
\caption{Right-moving electrons with Fermi energy $E_{\rm FR}=\mu_{\rm R}=0$ and left-moving electrons
with Fermi energy $E_{\rm FL}=\mu_{\rm L}=eV$ are backscattered off the time-dependent impurity $W(t)$
via $R^{\pm}$ and $L^{\pm}$ processes in which the electrons gain or lose an energy quantum $\hbar\omega$.}
\end{figure}

\begin{figure} \label{fig2}
  \hfill
  \includegraphics[width=4.9in]{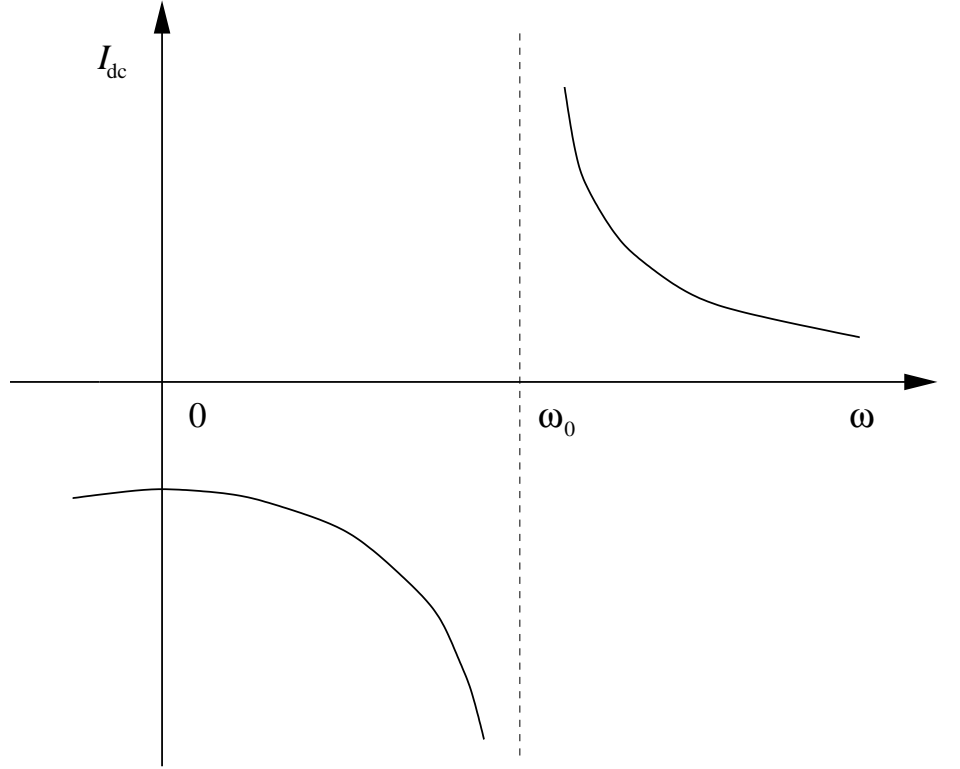}

  \hfill\hfill
\caption{Qualitative dependence of the backscattering contribution $I_{\rm dc}$ to the dc-current
on the frequency $\omega$ at $g<1/2$. $\omega_0=eV/\hbar$, where $V$ is the applied voltage.
Note that $I_{\rm dc}>0$
as $\omega>\omega_0$ (total current enhanced).
Our approach based on the lowest order of the perturbation theory is insufficient for
the region $\omega\approx\omega_0$.}
\end{figure}


\begin{references}
\bibitem{clear}
Note that when considering the product of the number of particles subject to backscattering
and the probability for such a process, it is not {\it a priori} clear whether it is the $L^+$
or $R^+$ that win.
\bibitem{foot}
As possible realization one may consider a polarized electron liquid or an edge state of
a quantum Hall bar. For electrons with spin the effective Luttinger parameter $g$ exceeds $1/2$
and our effect does not show up.
\bibitem{footvolt}
The electric potential $\phi$ in the point $x$ is proportional to
the energy cost of adding an infinitesimal charge to this point.
From the Hamiltonian (\ref{1}) one finds that $\phi\sim K(x)n(x)$,
where $K(x)$ is the interaction strength, $n(x)$ the charge
density. One sees that $\phi=0$ at $K(x)=0$ i.e. in the regions
where the electron-electron interaction is completely screened in
the left and right parts of the wire. Thus, the voltage drop
between these two region is always zero. This does not contradict
the fact that one can apply a finite voltage between the leads:
the voltage drop occurs in the contacts between the leads and the
wire. Note also that the electric potential is nonzero in the
region $|x|<L/2$ in the center of the wire. It depends on $x$ in a
non-monotonous way since $\phi=0$ at $|x|\gg L$.
\end{references}
\end{document}